\documentclass[tightenlines,aps,twocolumn,showpacs,nofootinbib]{revtex4}

\usepackage{float}
\usepackage{amsmath,amsfonts,graphicx,bm}

\newcommand{\be}{\begin{equation}}
\newcommand{\ee}{\end{equation}}
\newcommand{\ba}{\begin{eqnarray}}
\newcommand{\ea}{\end{eqnarray}}

\newcommand{\gw}{0.12\textwidth}

\begin{document}


%\preprint{ SLAC-PUB-???\\
% UH-????\\
% hep-ph/0507033}

\begin{titlepage}

\begin{flushright}
\vbox{
\begin{tabular}{l}
 \textsf{FERMILAB-PUB-05-317-T}\\
 \textsf{UH-511-1074-05}\\
 \textsf{hep-ph/0507033}
\end{tabular}
}
\end{flushright}

\title{One-loop matching of $\Delta S=2$ 
four-quark operators with improved staggered fermions}

\author{Thomas Becher} 
\affiliation{Fermi National Accelerator Laboratory,
P.~O.~Box 500, Batavia, IL 60510, USA}
\email{becher@fnal.gov}
\author{Elvira G\'amiz} 
\affiliation{Department of Physics and Astronomy, University of Glasgow,
Glasgow G12 8QQ, UK}
\email{e.gamiz@physics.gla.ac.uk}
\author{Kirill Melnikov}
\affiliation{Department of Physics and Astronomy, 
University of Hawaii, Honolulu, HI 96822, USA}
\email{kirill@phys.hawaii.edu}

\begin{abstract}
  We compute ${\cal O}(\alpha_s)$ lattice-to-continuum perturbative
  matching coefficients for the $\Delta S=2$ flavor changing
  four-quark operators for the Asqtad improved staggered
  action.  In conjunction with lattice simulations with three flavors
  of light, dynamical quarks, our results yield an unquenched
  determination of $B_K$, the parameter that determines the amount of
  indirect CP violation in the neutral kaon system. Its value is an
  important input for the unitarity triangle analysis of weak decays.
\end{abstract}

\maketitle

\thispagestyle{empty}
\end{titlepage}

\section{Introduction}
Lattice calculations have started extracting phenomenologically
relevant results with high precision and controlled uncertainties
\cite{Davies:2003ik}.  This requires simulations with dynamical
quarks, lattice actions with small discretization errors and good
control over perturbative corrections.  Effective field theory methods
are an important tool to achieve these goals. They enable a judicious
design of lattice actions with small discretization errors.  On the
other hand, the complicated form of improved lattice actions puts a
burden on perturbative calculations required to match the output of
lattice simulations to continuum physics.  For the matching to be
meaningful, perturbative calculations have to be performed with {\it
  exactly the same} action and operators, with which the
non-perturbative simulations are done.

The purpose of the present paper is to compute the matching
coefficients of the four-fermion $\Delta S=2$ operators relevant for
$K-\bar K$ mixing for the Asqtad lattice action, an improved staggered
fermion action used in many of the current simulations with dynamical
quarks. The strength of $K-\bar K$ mixing is determined by the matrix
element $\langle \bar K | (\bar s d)_{V-A} (\bar s d)_{V-A} | K
\rangle \sim B_K(\mu)$, where $\mu$ is the renormalization scale.  The
determination of $B_K(\mu)$ has become an important goal of lattice
simulations, since a precise value of this parameter translates into a
stringent constraint on the unitarity triangle.  The current
constraint is dominated by theoretical uncertainties, the largest
being the uncertainty in the value of $B_K$ \cite{Charles:2004jd}.
The experimental input comes from $K\rightarrow\pi\pi$ decays and has
percent level precision \cite{Eidelman:2004wy}.

Results of lattice calculations for $B_K(2~{\rm GeV})$ have been
summarized in \cite{Hashimoto:2004hn,Wingate:2004xa}; a central value
of $ B_K(2~{\rm GeV}) = 0.58(4)$ from quenched determinations was
advocated in \cite{Hashimoto:2004hn}. The uncertainty associated with
this value does not include the systematic error induced by quenching,
i.e.~neglecting the effects of dynamical fermions. The errors from
quenching are presumed to be of the order of 15\%, but are very hard
to estimate reliably \cite{Sharpe:1998hh}. In order to go beyond this
accuracy, the calculation needs to be performed with dynamical
fermions. The virtue of the improved action for which we perform our
calculation is that it allows for precise calculations with light,
dynamical quarks.  The action is used by the MILC collaboration
\cite{Bernard:2001av} and consists of the Asqtad discretization of
staggered quarks \cite{Lepage:1998vj,Orginos:1999cr} coupled to a
Symanzik improved gluon action \cite{Luscher:1985zq}.

The nonperturbative calculation of the matrix elements of the lattice
operators on the MILC dynamical configurations was discussed in
\cite{Gamiz:2004qx}, along with preliminary results for the bare $B_K$
parameter. Together with the matching coefficients given in this
paper, these results determine the renormalized continuum value
$B_K(\mu)$ \cite{CDGSW05}.

In \cite{Becher:2002if} two of us suggested an approach to lattice
perturbation theory that allows an efficient separation of energy
scales that appear in matching calculations -- the inverse lattice
spacing, which plays the role of the ultraviolet cut-off, and the
physical masses and momenta.  As a result of this scale separation, we
are able to treat the complicated integrals that appear in lattice
perturbation theory by algebraic means.  The approach allows a very
high degree of automation, thereby circumventing the complexity of
improved lattice actions to a certain degree. We have used this method
in \cite{Becher:2003fu} to compute the one-loop relation between the
bare lattice mass and the pole mass for the Asqtad action. In the
present paper we compute the matching coefficient for $\Delta S = 2$
four-fermion operators for the Asqtad action, using the algebraic
technique \cite{Becher:2002if} as well as numerically. The agreement
of the results obtained using these two completely different
techniques provides a strong check on the calculation.

The lattice-to-continuum matching coefficients for the $\Delta S=2$
four-quark operators have been previously calculated for the standard,
unimproved, discretization of staggered fermions
\cite{Ishizuka:1993fs}, as well as for a class of improved
staggered actions \cite{Lee:2003sk}. While none of these results is
directly relevant for the Asqtad action, we have checked that we
reproduce the known result for the unimproved action, once we switch off
all improvement terms.

The rest of the paper is organized as follows. In the next Section we
give the discretization of the four-fermion operators. In Section
\ref{sec:match} we discuss how the matching is performed and give the
result for each of the relevant Feynman diagrams. In Section
\ref{sec:TI} we include the tadpole improvement terms. We present our
results for the matching coefficients that relate bare lattice
operators to continuum ${\overline {\rm MS}}$ operators in Section
\ref{sec:results}.  Finally, in the last Section we summarize our
results and compare the size of the perturbative corrections found for
different staggered actions.

\section{Staggered and naive lattice operators}

For energy scales below the charm quark mass, the continuum effective 
Hamiltonian relevant for $K-\bar K$ mixing reads \cite{Buras:1990fn}
\begin{equation}\label{Heff}
   H_{\rm eff}^{\rm weak} = \frac{G_F^2 m_W^2}{16\pi^2} C(\mu)\, Q(\mu),
\end{equation}
where 
\begin{equation}
    Q = (\bar s d)_{V-A}\,(\bar s d)_{V-A}.
\label{qcont}
\end{equation}
The possibility to describe $K-\bar K$ mixing by a single operator is
a consequence of the $V-A$ structure of weak interactions and the
Fierz identities that facilitate expressing any relevant four-quark
operator in canonical form (\ref{qcont}).

Lattice simulations of the $K-\bar K$ mixing require introducing the
operator $Q$ on the lattice. It is convenient to perform the
perturbative calculations with naive instead of staggered quarks and
to carry out the spin-diagonalization and the reduction from the sixteen
to four doublers (or ``tastes'') afterwards.  We use the standard
discretization for the staggered four-quark operators and now show how
these operators can be rewritten in terms of the naive fermion field.

The naive fermion field $\psi$ and the staggered field
$\chi$ at lattice site $n=(n_1,n_2,n_3,n_4)$ are related by
\begin{align}
\psi(n)&=\Gamma_n \,\chi(n) &
\bar\psi(n)&=\bar\chi(n)\,\Gamma_n^\dagger \label{eq:psi}
\end{align}
where
\begin{equation}
\Gamma_n=(\gamma_1)^{n_1}\,(\gamma_2)^{n_2}\,(\gamma_3)^{n_3}\,(\gamma_4)^{n_4} \,.
\end{equation}
We choose a Hermitian representation of the Euclidean Dirac algebra with
\begin{align}
\left \{ \gamma_i, \gamma_j \right\}&=2\delta_{ij} & &\text{and} &\gamma_5&=\gamma_1\,\gamma_2\,\gamma_3\,\gamma_4 \,.
\end{align}

When written in terms of staggered fermions, the quark action becomes
spin diagonal and three of the four components of the quark field
$\chi$ can be dropped. In perturbation theory it is more convenient to
keep all components, and replace $n_f\rightarrow n_f/4$ at the end
of the calculations to obtain the result for the single component
field. 

The fields $\chi(2N+A)$ are collected into a set of Dirac fields
$q(2N)$ that live on the even lattice sites and are spread over a unit
hypercube \cite{Gliozzi:1982ib, Kluberg-Stern:1983dg,Sharpe:1993ur} 
($A_\mu=0$ or 1, $\mu=1\dots 4$)
\begin{equation}
q(2N)_{\alpha i j}=\frac{1}{8}\sum_A (\Gamma_A)_{\alpha,i}\; \chi_j(2N+A)\,.
\end{equation}
The index $\alpha$ is the Dirac index of the new field and $i$ the
``taste'' index. The second taste index $j$ runs over the components
of the field $\chi$. Staggered fermions have only one component, while
the four component of the naive field give rise to $4\times4$
tastes. Bilinear quark operators with spin structure $\gamma_S$ and taste structure $\xi_T=\Gamma_T^*$ (with $T$ again on the unit hypercube) are
\begin{multline}
\bar q(2N) (\gamma_S\otimes \xi_T) q(2N)\\=\frac{1}{16}\,\sum_{A,B} \bar\chi(2N+A) \;\chi(2N+B)\, \frac{1}{4}{\rm tr}( \Gamma_A^\dagger\,\gamma_S \Gamma_B\,\Gamma_T^\dagger ) \,.
\end{multline}
Since the quark fields in the operators are at different lattice
points, they need to be connected by Wilson lines in order to make the
operators gauge invariant. We suppress these Wilson lines, but it is
understood that gauge strings among all possible shortest paths
connecting the quark fields are inserted and the operator is divided
by the number of paths.  Using (\ref{eq:psi}) we rewrite the operator
in terms of the naive fermion field:
\begin{multline}
\bar q(2N) (\gamma_S\otimes \xi_T) q(2N) \\
=\frac{1}{16}\,\sum_{A,B} \bar\psi(2N+A)\,\Gamma_A\, \Gamma_B^\dagger\,\psi(2N+B)\, \frac{1}{4}{\rm tr}( \Gamma_A^\dagger\,\gamma_S \Gamma_B\,\gamma_T^\dagger )\,.
\end{multline}
 Let us write out the vector and axial currents at $N=0$ with unit taste structure in
terms of the naive quark field $\psi(n)$. We find
\begin{multline}
 \bar q\, (\gamma_\mu\otimes 1)\, q
= \frac{1}{16}\sum_{A} \delta_{A_\mu,0} \\ \left[
  \bar\psi(A)\gamma_\mu\psi(A+\hat\mu)+\bar\psi(A+\hat\mu)\gamma_\mu\psi(A)\right]\,, \label{eq:Amu}
\end{multline}
and ($\overline{A}=(1,1,1,1)-A$)
\begin{multline}
 \bar q\, (\gamma_\mu\gamma_5\otimes 1)\, q 
= \frac{1}{16}\sum_{A} \delta_{A_\mu,0} \\ \left[
  \bar\psi(A)\gamma_\mu\gamma_5\psi(\overline{A+\hat\mu})+\bar\psi(\overline{A})\gamma_\mu\gamma_5\psi(A+\hat\mu)\right] \,.\label{eq:Vmu}
\end{multline}
In the continuum limit, the massless staggered action has a
$SU(4)_L\times SU(4)_R$ chiral symmetry (for a single staggered field)
and one would naively expect to find fifteen Goldstone bosons after
chiral symmetry breaking, corresponding to the fifteen traceless taste
matrices.  However, this symmetry is explicitly broken to an axial
$U(1)$ symmetry by terms in the action proportional to the lattice
spacing squared.  The generator of the remaining axial symmetry is the
matrix $\gamma_S\otimes\xi_T=\gamma_5\otimes\gamma_5$. In our case,
this implies that only the kaon with taste structure $\gamma_5$
becomes massless in the chiral limit. For this reason, the simulation
is done with currents having $\xi_T=\gamma_5$ taste structure and not
the unit matrix as in (\ref{eq:Amu}) and (\ref{eq:Vmu}).  However, it
has been shown that the result for diagrams in which the fermions in
the operators are not contracted are identical for the Dirac and taste
structures $\Gamma\otimes \Gamma'$ and $\Gamma\gamma_5\otimes
\Gamma'\gamma_5$ \cite{Sharpe:1993ur}.\footnote{The reason is that in
  this case, one can view the four fermions as four different quark
  flavors and perform separate axial $U(1)$ transformations on each
  field.} For our perturbative calculation we will thus use
(\ref{eq:Vmu}) instead of the vector current $\gamma_\mu\otimes
\gamma_5$ and (\ref{eq:Amu}) in place of $\gamma_\mu\gamma_5\otimes
\gamma_5$.

The operator $Q$ for which we want to perform the matching calculation
is a product of two $V-A$ currents. We will see that QCD corrections
to bare lattice four-quark operators affect the vector and axial parts
differently; as a consequence currents of the form $V+A$ are
generated. A minimal set of lattice operators that matches to the
continuum and closes under renormalization consists of 4 scalar and 2
pseudoscalar operators. Schematically these operators are
\begin{align}
&\begin{aligned}  
Q_{1} &= (\bar s^a d^a)_V (\bar s^b d^b)_V, &
\phantom{o}& Q_{2}= (\bar s^a d^b)_V (\bar s^b d^a)_V, \nonumber\\
 Q_{3} &= (\bar s^a d^a)_A (\bar s^b d^b)_A, & 
\phantom{o}& Q_4 = (\bar s^a d^b)_A (\bar s^b d^a)_A,  
\end{aligned} \label{ques}\\
&\begin{aligned}  
& Q_{5} = (\bar s^a d^a)_V (\bar s^b d^b)_A+(\bar s^a d^a)_A (\bar s^b
d^b)_V, \\
& Q_{6}= (\bar s^a d^b)_V (\bar s^b d^a)_A+(\bar s^a d^b)_A (\bar s^b d^a)_V\,.
\end{aligned}
\end{align}
$V$ and $A$ are the vector and axial currents with taste structure
$\xi_T=\gamma_5$. The color indices $a,b$ indicate which fields are
connected by Wilson lines. The pseudoscalar operators $Q_{5,6}$ only
mix amongst themselves and their $K-\bar K$ matrix element vanishes,
because of parity. In the following we will therefore only consider
the renormalization of the operators $Q_{1-4}$.

The above operators are used in the nonperturbative calculation of the
matrix elements by the HPQCD collaboration \cite{Gamiz:2004qx,
  CDGSW05}. In contrast to the lattice action, these operators are not
Symanzik improved.

\section{Matching calculation\label{sec:match}}

To perform the matching we calculate a physical quantity in the
continuum and on the lattice, expand around the continuum limit, and
adjust the Wilson coefficients of the lattice operators in such a way
that they reproduce the continuum result. Specifically, we determine
the bare lattice Wilson coefficients from the quark-quark scattering
amplitude
\begin{align}
{\cal A}= Z^2 C^{\rm bare}_{i} \langle O^{\rm bare}_i \rangle= Z^2 C_{i}\,Z_{ij} \langle O_j \rangle = 
Z_{\rm lat}^2 C^{\rm lat}_{i} \langle O^{\rm lat}_i \rangle\, ,\label{eq:matching}
\end{align}
where $Z$ and $Z_{\rm lat}$ are on-shell quark wave-function
renormalizations and $\langle O \rangle$ is the amputated four-quark
Green's function with an insertion of the operator $O$. The
lattice-to-continuum matching coefficients are independent of the
quantity chosen to perform the matching. At one-loop order, it is
simplest to regulate infrared divergencies with a gluon mass and to
calculate the scattering amplitude of massless quarks at zero external
momentum.

At tree level in the continuum limit, the operators $Q_i$ are
identical to their continuum counterparts; this implies that no tree
level matching is required.  At one loop the matching coefficients
$C_i^{\rm lat}$ are obtained by calculating the difference between the
lattice and continuum one-loop diagrams. In the difference, the
infrared divergencies cancel. Below we present such a calculation for
the set of lattice operators described above.

For unimproved staggered fermions the renormalization of the axial
current operator with
$\gamma_S\otimes\xi_T=\gamma_\mu\gamma_5\otimes\gamma_5$ is finite and
identical in the continuum and on the lattice. This happens because it
renormalizes in the same way as the vector current with unit taste
structure $\gamma_\mu\otimes 1$ which has exactly the same form as the
quark-gluon coupling in the Lagrangian.  Because of the improvement
terms, this is not true for the Asqtad action and we will need to
evaluate the matching not only for the four-quark operators but also
for the current.

The matching calculation requires computing matrix elements 
of certain operators in lattice perturbation theory; to do so
 we 
use the approach described in \cite{Becher:2002if,Becher:2003fu}.
In those  references an expansion of the lattice integrals 
around the continuum limit is introduced by utilizing the 
technique of asymptotic expansions.
The asymptotic expansion  around the continuum limit
splits the lattice diagrams into two parts: (i) a soft part which is
obtained by evaluating continuum loop integrals (in analytic regularization)
and is independent of
the discretization and (ii) a hard part which depends on the
discretization but is independent of particle masses and momenta.
The proximity of the soft part of the lattice loop integrals 
and the continuum integrals in dimensional regularization 
permits significant simplifications when computing the difference
of the lattice and continuum Green's functions required for 
matching calculations. 

There are five diagrams (plus permutations) that contribute at ${\cal
  O}(\alpha_s)$ to the quark-quark scattering amplitude; they are
shown in Fig.~\ref{fig:graphs}.  To present the result of the
calculations, we introduce the following notation for the difference
of the lattice and continuum one-loop matrix elements of the
four-fermion operators: \be \delta \langle Q^{i} \rangle_\alpha =
\left ( \frac{\alpha_s}{4\pi} \right ) \left (\frac{\mu a}{2} \right
)^{2\epsilon} \sum \limits_{j=1}^4 q_{\alpha,j}^{i} Q^{j}, \ee where
the label $\alpha$ refers to a particular diagram shown in
Fig.~\ref{fig:graphs}.  The spacetime dimension is $d=4-2\epsilon$,
$\mu$ is the renormalization scale in the ${\overline {\rm MS}}$
subtraction scheme and $a$ the lattice spacing. In what follows, we
split the matrices $q_{i}$ into soft and hard pieces. We give the
result for each diagram separately. We hope that the results for the
individual diagrams will be useful for readers performing similar
calculations in the future.

\subsection{Soft part}
\begin{figure}
\begin{center}
\begin{tabular}{ccc}
\includegraphics[width=\gw]{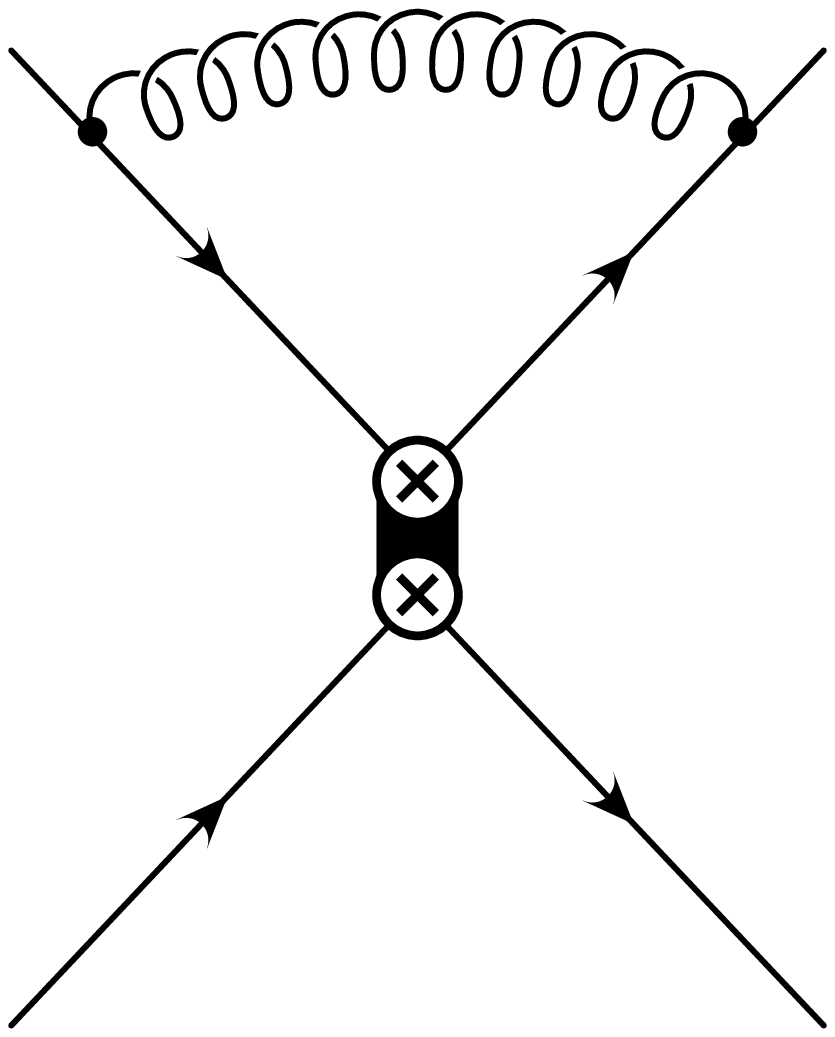} &
\includegraphics[width=\gw]{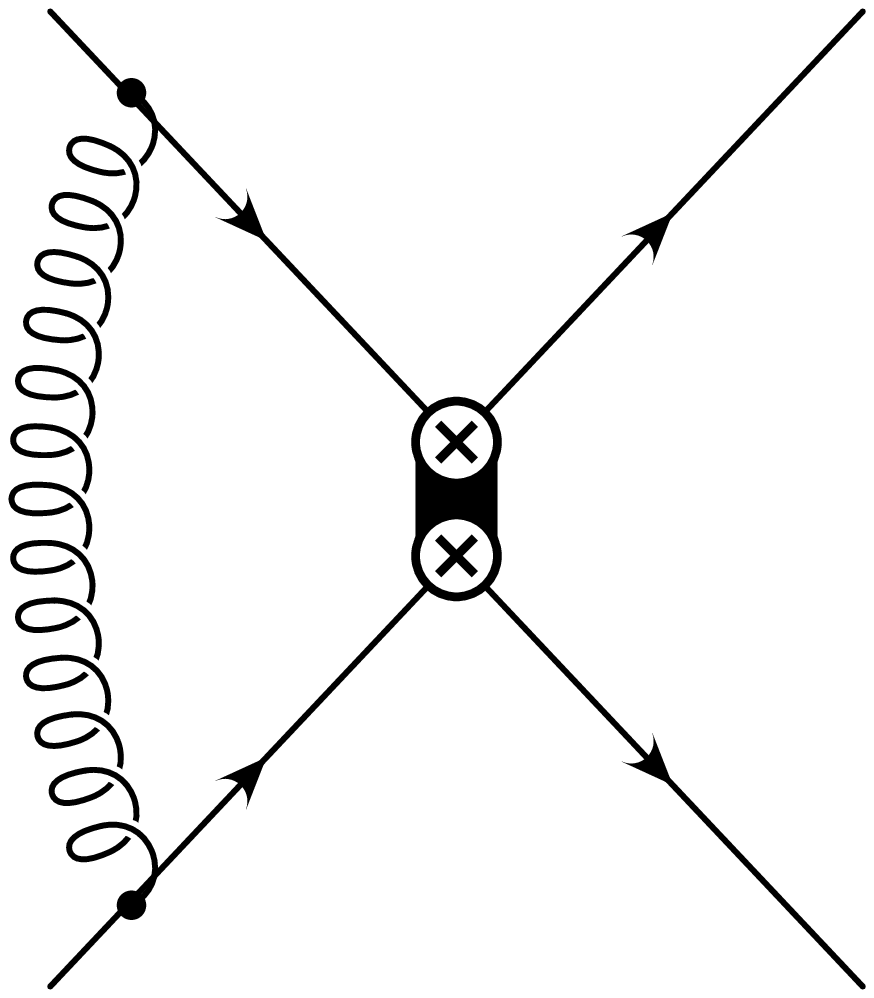} &
\includegraphics[width=\gw]{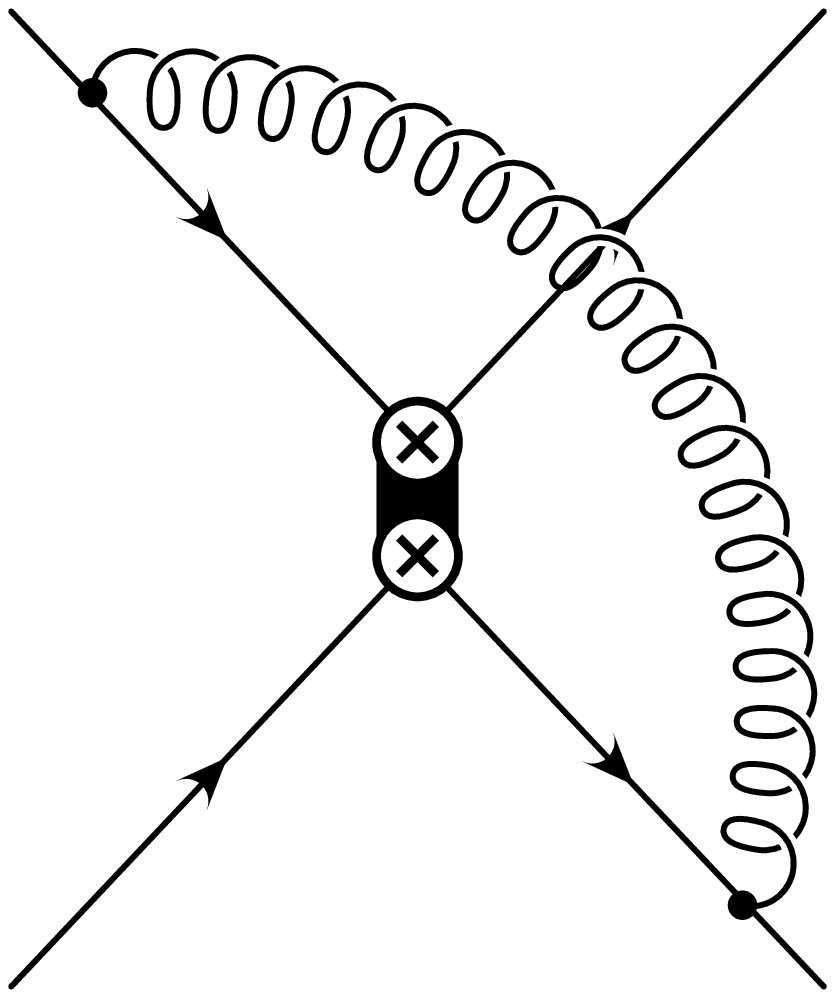} 
\end{tabular}
\begin{tabular}{cccc}
\includegraphics[width=\gw]{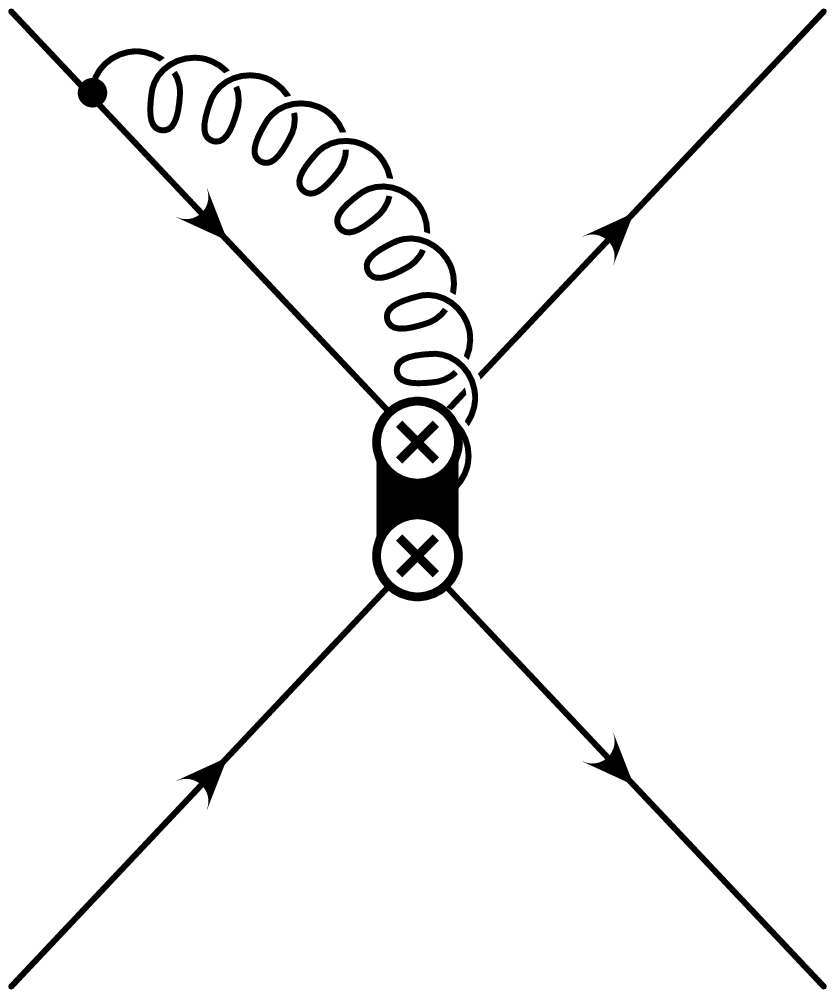} &
\includegraphics[width=\gw]{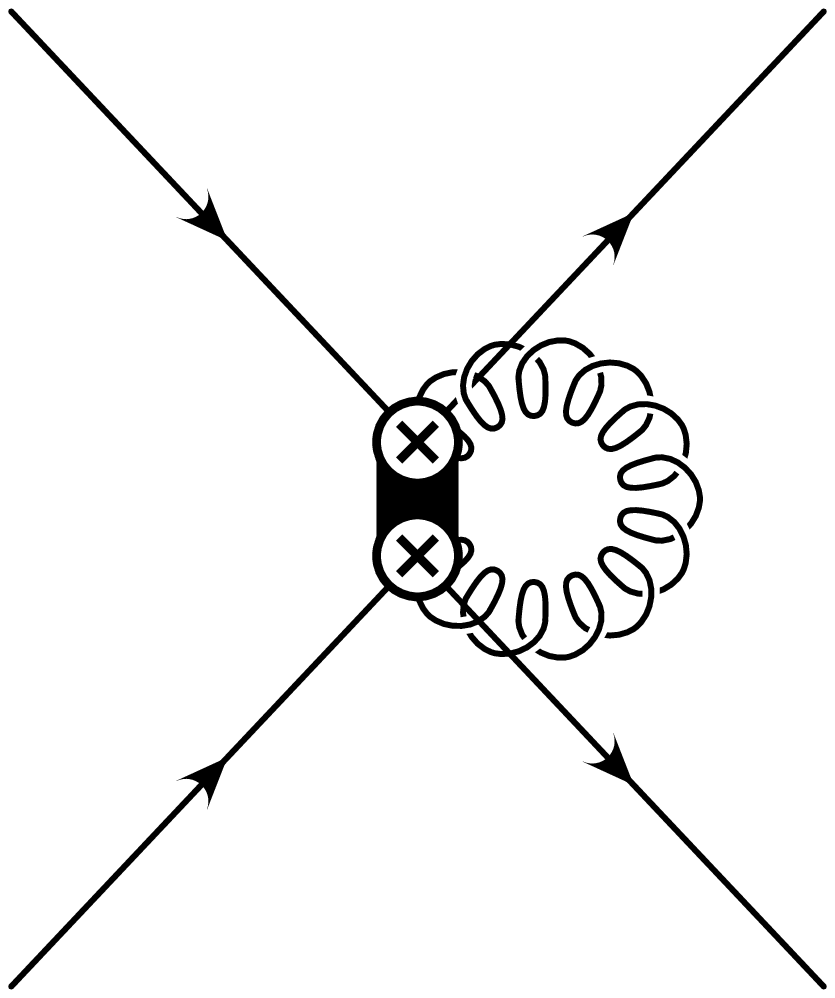} &
\end{tabular}
\end{center}
\caption{Current-current diagrams. The diagrams on the second line do not have a continuum analog. Not shown are additional diagrams that can be obtained by flipping one of the above diagrams horizontally or vertically.}\label{fig:graphs}
\end{figure}

 As a first step in the matching calculation, we compute  the
difference between the soft parts of lattice matrix elements 
and the  result derived  in
dimensional regularization. To obtain the full result, we later add
the hard part for a given lattice action. Technically the soft part is
obtained by evaluating the continuum diagrams shown in Fig.~\ref{fig:graphs},  
first in dimensional and then in analytic
regularization. The calculation is simplified by noting that a
difference between the two regulators can only arise from
ultraviolet divergent loop integrals.

Below we give the results for the {\em difference} of the
soft parts of the matrices $q$ and the corresponding 
continuum diagrams. Since the  
diagrams 4 and 5 in Fig.~\ref{fig:graphs}  do not have continuum analog, their 
contribution to the soft part is zero: 
$q_4^{\rm soft}=q_5^{\rm soft}= 0$.
For diagram $1$, we derive
\begin{equation*}
q_{1}^{\rm soft} = \left (
\begin{array}{cccc}
C_F f_{11}^{(1)} & 0 & 0 & 0 \\
\frac{f_{11}^{(1)}}{2} & -\frac{f_{11}^{(1)}}{2N} & 0 & 0 \\
0 & 0 & C_F f_{11}^{(1)} & 0 \\
0 & 0 & \frac{f_{11}^{(1)}}{2} & -\frac{f_{11}^{(1)}}{2N}   \\
\end{array}
\right )\,.
\end{equation*}
The variable $N=3$ denotes the number of colors, $C_F=(N^2-1)/2N$
and
\begin{eqnarray}
&& f_{11}^{(1)} = -2\,(1+\xi)\,\left(\frac{1}{\epsilon } -
\frac{1}{\delta}\right) +1 - 2\xi\,.  
\end{eqnarray}
The parameter $\delta$ is an intermediate analytic regulator, which
will drop out in the sum of the hard and soft part
\cite{Becher:2002if}. Setting the gauge parameter $\xi=0$, one obtains
the result in Feynman gauge while $\xi=-1$ corresponds to Landau gauge. For
details on our notation and method of calculation, we refer the reader
to \cite{Becher:2003fu}.

For diagrams 2 and 3 in Fig.~\ref{fig:graphs} we obtain,
\begin{equation*}
q_{2}^{\rm soft} = \left (
\begin{array}{cccc}
-\frac{f_{11}^{(2)}}{2N}  & \frac{f_{11}^{(2)} }{2} 
& -\frac{f_{13}^{(2)}}{2N}  & \frac{f_{13}^{(2)}}{2} \\
\frac{ f_{11}^{(2)}}{2} & -\frac{f_{11}^{(2)}}{2N}
& \frac{f_{13}^{(2)}}{2}  & -\frac{f_{13}^{(2)}}{2N}  \\
 -\frac{f_{13}^{(2)}}{2N} & \frac{f_{13}^{(2)}}{2} 
&  -\frac{f_{11}^{(2)}}{2N} & \frac{f_{11}^{(2)}}{2} \\
\frac{f_{13}^{(2)}}{2} & -\frac{f_{13}^{(2)}}{2N} 
&  \frac{f_{11}^{(2)}}{2} & -\frac{f_{11}^{(2)}}{2N} 
\end{array}
\right ),
\end{equation*}
\begin{equation*}
q_{3}^{\rm soft} = \left (
\begin{array}{cccc}
-\frac{f_{11}^{(3)}}{2N}  & \frac{f_{11}^{(3)} }{2} 
& -\frac{f_{13}^{(3)}}{2N}  & \frac{f_{13}^{(3)}}{2} \\
0 & C_F f_{11}^{(3)}
& 0  & C_F f_{13}^{(3)}   \\
 -\frac{f_{13}^{(3)}}{2N} & \frac{f_{13}^{(3)}}{2} 
&  -\frac{f_{11}^{(3)}}{2N} & \frac{f_{11}^{(3)}}{2}  \\
0 & C_F f_{13}^{(3)} 
&  0 & C_F f_{11}^{(3)} 
\end{array}
\right ), 
\end{equation*}
where
\begin{eqnarray} 
&& f_{11}^{(2)} = -f_{11}^{(3)}+6=\left(5+2\xi \right )\left(
\frac{1}{\epsilon}-\frac{1}{\delta} \right ) +\frac{15}{2} + 2\xi,
\nonumber \\ 
&& f_{13}^{(2)} =f_{13}^{(3)}=3\left
(\frac{1}{\epsilon}-\frac{1}{\delta} \right ) +\frac{5}{2}.
\end{eqnarray}
To evaluate the continuum diagrams, a prescription for the treatment
of $\gamma_5$ and the extension of the Fierz identities to
$d$-dimensions \cite{Dugan:1990df,Buras:1989xd,Herrlich:1994kh} has to
be adopted. We use naive dimensional regularization for $\gamma_5$ and
treat the evanescent operators exactly as in \cite{Buras:1990fn}, to
be compatible with the results for the Wilson coefficients given in
these references. Note that with this prescription, the operators
$Q_1\propto\gamma_\mu\otimes\gamma_\mu$ and
$Q_3\propto\gamma_\mu\gamma_5\otimes\gamma_\mu\gamma_5$ also mix into
operators with Dirac structure $1\otimes1$, $\gamma_5\otimes\gamma_5$
and $\sigma_{\mu\nu}\otimes\sigma_{\mu\nu}$. We do not give the result
for this mixing, since it drops out in the sums $Q_1+Q_3$ and
$Q_2+Q_4$ which are relevant for our matching calculation. However,
separating the vector and axial contributions also in the soft part
allows us to read off the current renormalizations from the first
diagram in Fig.~\ref{fig:graphs} and provides additional consistency
checks, such as cancellation of $\delta$-poles and gauge invariance of
the sum of hard and soft parts.

In addition to the diagrams shown in Fig.~\ref{fig:graphs}, we have
to account for the external leg corrections, given by the wave
function renormalization of the massless quarks. This correction is
 universal in that it does not depend on the operator under
consideration.  We obtain 
\be 
{\delta Z}^{(1)} O_i|_{\rm soft} =C_F \left(
\frac{2}{\epsilon }- \frac{2}{\delta} +1 \right) (1+\xi) O_i, 
\ee
where $Z^2_{\rm lat} - Z^2$=$\frac{\alpha_s}{\pi}\,\delta Z^{(1)}+\dots$ 
is the difference between the lattice (soft part only) and the on-shell 
quark wave function renormalization constant. This difference
vanishes in Landau gauge, which is a consequence of the
fact that in this gauge the continuum quark wave function
renormalization is finite.

Note that the coefficients of the $1/\epsilon$ and $1/\delta$ pieces
in all the equations above
are exactly opposite. The $1/\delta$-divergences will cancel against
those of the hard parts. The fact that the above differences do not
depend on any infrared scales (the gluon mass in our case) shows
that the result of a matching calculation is process independent.

\subsection{Hard part}

We now give the result of our calculation of the hard part. 
This amounts to presenting the matrices $q_{\alpha}$
for the five diagrams in Fig.~\ref{fig:graphs}. 
The symmetry factors for each of the 
diagrams are included. We begin with naive unimproved 
fermions coupled to unimproved glue.

The result for the hard part of the first three diagrams in
Fig.~\ref{fig:graphs}, has the same structure as was found for
the soft part: 

\begin{align*} q_{1}^{\rm hard} &= \left (
\begin{array}{cccc}
C_F d_{11}^{(1)} & 0 & 0 & 0  \\
\frac{d_{11}^{(1)}}{2} & -\frac{d_{11}^{(1)}}{2N} & 0 & 0 \\
0 & 0 & C_F d_{33}^{(1)} & 0 \\
0 & 0 & \frac{d_{33}^{(1)}}{2} & -\frac{d_{33}^{(1)}}{2N}   
\end{array}
\right )\,, 
\end{align*}
where
\begin{align}
 d_{11}^{(1)} &= 
- \frac{2(1+\xi)}{\delta} -3.53 - 2.38\,\xi, \nonumber \\ 
 d_{33}^{(1)} &= 
- \frac{2(1+\xi)}{\delta}  -3.62 + 0.695\,\xi \,,
\end{align}
and
\begin{align*}
q_{2}^{\rm hard} &= \left (
\begin{array}{cccc}
-\frac{d_{11}^{(2)}}{2N}  & \frac{d_{11}^{(2)} }{2} 
& -\frac{d_{13}^{(2)}}{2N}  & \frac{d_{13}^{(2)}}{2}  \\
\frac{ d_{11}^{(2)}}{2} & -\frac{d_{11}^{(2)}}{2N}
& \frac{d_{13}^{(2)}}{2}  & -\frac{d_{13}^{(2)}}{2N}\\
 -\frac{d_{13}^{(2)}}{2N} & \frac{d_{13}^{(2)}}{2} 
&  -\frac{d_{11}^{(2)}}{2N} & \frac{d_{11}^{(2)}}{2} \\
\frac{d_{13}^{(2)}}{2} & -\frac{d_{13}^{(2)}}{2N} 
&  \frac{d_{11}^{(2)}}{2} & -\frac{d_{11}^{(2)}}{2N} 
\end{array}
\right ), \,
 \nonumber\\ &\nonumber \\ 
q_{3}^{\rm hard} &= \left (
\begin{array}{cccc}
-\frac{d_{11}^{(3)}}{2N}  & \frac{d_{11}^{(3)} }{2} 
& -\frac{d_{13}^{(3)}}{2N}  & \frac{d_{13}^{(3)}}{2}   \\
0 & C_F d_{11}^{(3)}
& 0  & C_F d_{13}^{(3)}  \\
 -\frac{d_{13}^{(3)}}{2N} & \frac{d_{13}^{(3)}}{2} 
&  -\frac{d_{11}^{(3)}}{2N} & \frac{d_{11}^{(3)}}{2}  \\
0 & C_F d_{13}^{(3)} &  0 & C_F d_{11}^{(3)}  
\end{array}
\right ),
\end{align*}
with
\begin{align}
 d_{11}^{(2)} &=-d_{11}^{(3)} = 
\frac{(5+2\xi)}{\delta} +0.157 + 0.689\xi, \nonumber \\ 
 d_{13}^{(2)} &=d_{13}^{(3)} = 
\frac{3}{\delta} -0.341,
\end{align}

The remaining two diagrams have no continuum analogue. We obtain
\begin{align*}
q_{i=4,5}^{\rm hard} &=\left (
\begin{array}{cccc}
C_F\,d_{11}^{(i)} & 0  & 0 & 0\\
d_{21}^{(i)} & \!\!\!C_F\,d_{22}^{(i)}-\frac{d_{21}^{(i)}}{N} \!\!\! & 0 & 0  \\
0 & 0 &   C_F\,d_{33}^{(i)} & 0  \\
0 & 0 & d_{43}^{(i)} &\!\!\!C_F\,d_{44}^{(i)}-\frac{d_{43}^{(i)}}{N}\!\!\! \\
\end{array}
\right )\,, 
\end{align*}
with entries
\begin{align}
d_{11}^{(4)} &= 16.06+18.38\xi, \nonumber \\ 
d_{21}^{(4)} &= 1.18 + 1.69\xi, \nonumber\\
d_{22}^{(4)} &= 13.70 + 15.00\xi,\\ 
d_{33}^{(4)} &= 12.23+12.23\xi, \nonumber \\
d_{43}^{(4)} &= -0.73-1.38\xi, \nonumber \\
d_{44}^{(4)} &= 13.70 + 15.00\xi, \nonumber
\end{align}
\begin{align}
d_{11}^{(5)} &= -73.40-9.19\,\xi ,  \nonumber\\
d_{21}^{(5)} &= -2.20-0.85\xi, \nonumber\\
d_{22}^{(5)} &= -48.9-7.50\xi, \\
d_{33}^{(5)} &= -24.47 - 6.12\xi, \nonumber\\  
d_{43}^{(5)} &=1.57+0.69\xi, \nonumber \\
d_{44}^{(5)} &= -48.9-7.50\xi. \nonumber 
\end{align}
We also give the result for the hard part 
fermion wave function renormalization on the lattice:
\ba
{\delta Z}^{(1)}\, O_{i}|_{{\rm hard}} && =C_F  
\Big (
-10.614 -11.928\,\xi 
+ \frac{2(1+\xi)}{\delta} 
\nonumber \\
&& 
+\left[24.466 + 6.117\,\xi  \right] \Big )\, O_i. 
\label{zhardN}
\ea
We have separated out the tadpole
contribution in (\ref{zhardN}) in square brackets. In Landau gauge 
this contribution exactly cancels against the tadpole improvement
term. 

We now give the results for the entries of the matrices for the 
case of Asqtad action. 
In this case, as explained in \cite{Becher:2003fu},
we expand 
in the improvement terms, for both quark and gluon actions.  
We use half of the highest order terms as the numerical error estimate 
and multiply the last term of the series by $2/3$ to resum the higher
order contributions that behave like an alternating geometric series
with expansion parameter $1/2$ (see \cite{Becher:2003fu} for details).
We find 
\begin{align}
 d_{11}^{(1)} &= 
- \frac{2(1+\xi)}{\delta} -3.00(1) - 2.379\,\xi, \nonumber \\ 
 d_{33}^{(1)} &= 
- \frac{2(1+\xi)}{\delta} + 0.81(4)  + 0.695\,\xi \nonumber \\
%&\nonumber\\
%\end{align} 
%\begin{align}
 d_{11}^{(2)} &=-d_{11}^{(3)}= \frac{(5+2\xi)}{\delta} +2.79(10)+0.689\xi, \nonumber \\
 d_{13}^{(2)} &=d_{13}^{(3)}= \frac{3}{\delta}+2.10(9), \nonumber \\ &\nonumber\\
%\end{align}
%and
%\begin{align}
 d_{11}^{(4)} &= 18.54(6) +18.38\xi, \nonumber \\
d_{21}^{(4)} &=1.73(1)+1.69 \xi, \nonumber\\
d_{22}^{(4)} &=15.09(3)+15.00\xi, \\ 
d_{33}^{(4)} &=12.23+12.23 \xi, \nonumber \\
d_{43}^{(4)} &=-1.43(2)-1.38\xi, \nonumber \\
d_{44}^{(4)} &=15.09(3)+15.00 \xi, \nonumber \\ &\nonumber\\
 d_{11}^{(5)} &= -58.0(5)-9.19\,\xi , \nonumber \\
 d_{21}^{(5)} &= -2.06-0.84\,\xi , \nonumber \\
 d_{21}^{(5)} &= -39.1(3)-7.50\,\xi , \nonumber \\
 d_{33}^{(5)} &= -20.3(1) - 6.12\,\xi ,\nonumber \\ 
 d_{43}^{(5)} &= 1.56(1)+0.69\,\xi , \nonumber \\
 d_{44}^{(5)} &= -39.1(3)-7.50\,\xi . \nonumber 
\end{align}
The gauge dependent part is the same as in the unimproved case because
the improvement terms in the action are transverse \cite{Becher:2003fu}.

For the Asqtad action, the wave function renormalization 
contribution is:
\ba
{\delta Z}^{(1)}\, O_{i}|_{{\rm hard}} && =C_F  
\Big (
-14.3(2) -11.28\,\xi
+ \frac{2(1+\xi)}{\delta} 
\nonumber \\
&& 
+\left[ 30.1(6) 
+ 5.47 \xi \right] \Big )\,O_i.
\label{zhard}
\ea

The renormalization of the axial and vector current can be read off
from the above expressions. In both cases it is given by one half of
the corresponding entries in the matrices $q_i$ for diagrams 1, 4 and
5 in Fig.~\ref{fig:graphs} and the contribution due to the wave
function renormalization.  As discussed earlier in this section,
the discretized version of the axial current that we use in this paper
renormalizes identically to the unimproved quark-gluon vertex.  As a
consequence, the lattice-to-continuum matching coefficient for the
axial current should be equal to unity for the unimproved action, to
all orders in $\alpha_s$. Using the results presented above, it is
easy to verify this at the one-loop level.

\subsection{Matching coefficients from a different method\label{sec:method2}}

Most of the diagrams needed in the determination of the matching
coefficients for four-fermion operators can be calculated from the
renormalization coefficients for bilinears operators, as was first
pointed out by Martinelli \cite{Martinelli84}. The method has been
generalized to Landau-gauge operators in \cite{Sharpe:1993ur}, general
local operators in \cite{GBS96} and staggered gauge invariant
operators in \cite{Lee:2003sk}.

\begin{figure}
\begin{center}
\begin{tabular}{ccccc}
\includegraphics[width=\gw,angle=180]{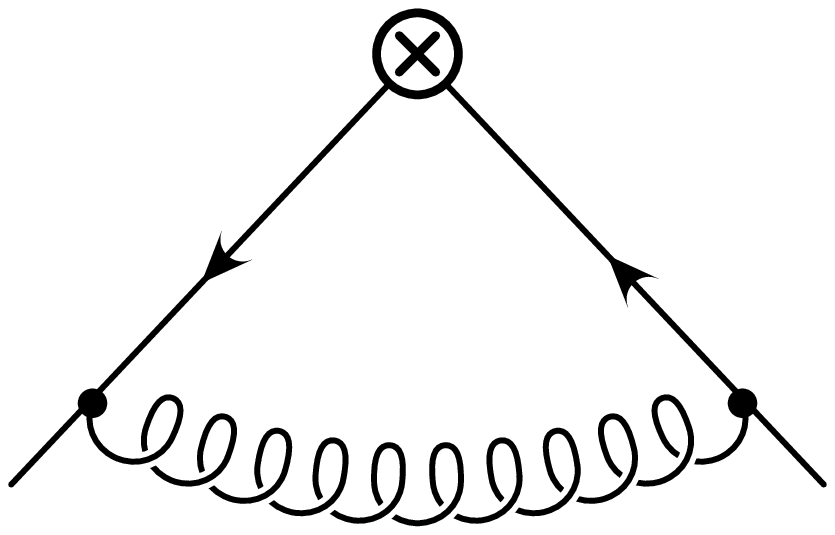} &&
\includegraphics[width=\gw,angle=180]{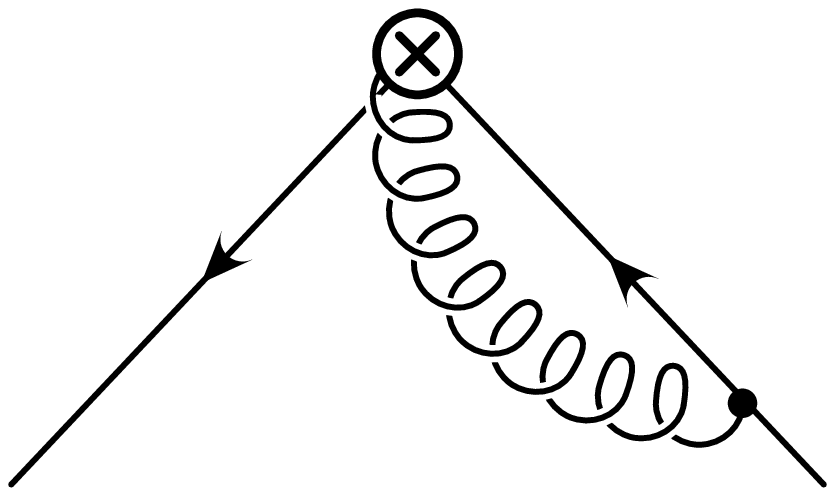} && 
\raisebox{0.027\textwidth}
{\includegraphics[width=\gw,angle=180]{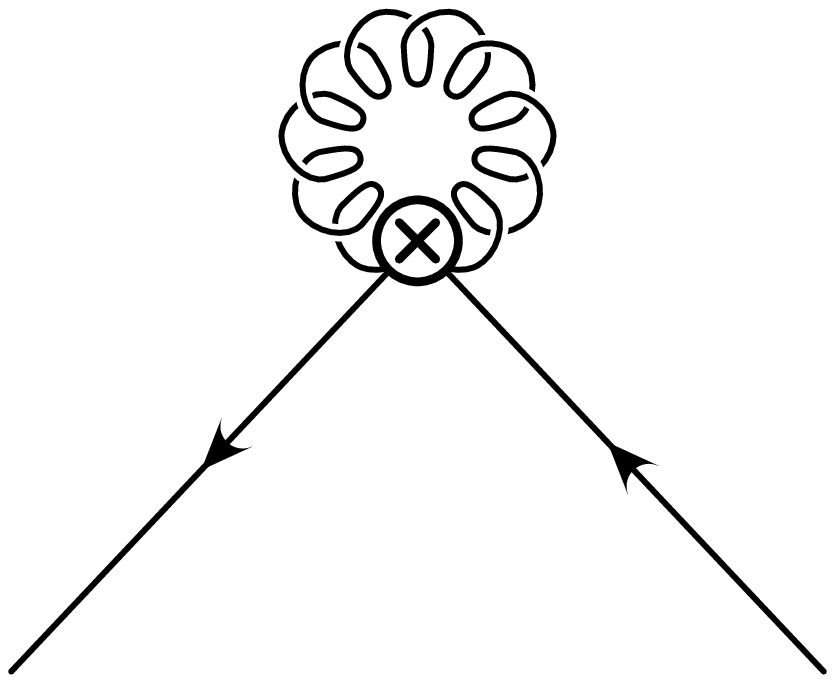}}
\end{tabular}
\end{center}
\caption{Current diagrams.}\label{fig:bilgraphs}
\end{figure}

In Fig.~\ref{fig:graphs}, we did not show mirrored diagrams and did
not display the color structure of the four-quark operators. Also, we
have not indicated from which of the two currents the gluons in
diagrams 4 and 5 are emitted. Listing all possibilities, one ends up
with 16 diagrams of type 4 and 6 of type 5. Only 8 of the 16 diagrams
of type 4 and 2 of the 6 diagrams of type 5 are genuine four-fermion
contributions. All other diagrams factor into a product of a one-loop
current renormalization diagram times a tree level current.
Furthermore, only half of the remaining diagrams contribute to the
mixing between operators whose bilinears have identical
taste-structure, which is all that is relevant for matrix elements
involving external states of definite taste. One cannot avoid
calculating the genuine four-fermion diagrams, but the result for the
other topologies can be obtained from the diagonal renormalization
coefficients for bilinears operators using Fierz and charge
conjugation transformations as described in \cite{Lee:2003sk}.  Apart
from the 5 genuine four-fermion diagrams, the calculation reduces to
the evaluation of the current renormalization diagrams in
Fig.~\ref{fig:bilgraphs} and the universal wave function
renormalization of the massless quarks. In Fig.~\ref{fig:bilgraphs},
the crossed circle denotes the insertion of any bilinear with structure
$(\gamma_S\otimes\xi_T)$. The calculation is further simplified by
noting that the corrections are identical if the operators are multiplied
by $(\gamma_5\otimes\gamma_5)$ due to the axial symmetry, as stressed
above.

For diagonal four-fermion operators (with identical taste and spin
structure in both bilinears) this procedure gives us a $35\times35$
matrix from which we can extract the $2\times2$ block describing the
mixing between $[\gamma_{\mu}\otimes\gamma_5]
[\gamma_{\mu}\otimes\gamma_5]$ and
$[\gamma_{\mu}\gamma_5\otimes\gamma_5]
[\gamma_{\mu}\gamma_5\otimes\gamma_5]$. This block is related to the
$4\times4$ matrix in (\ref{match}) by including the appropriate color
factors as explained in \cite{Lee:2003sk}. 

We have done the calculation of the matching coefficients for the
$\Delta S=2$ four-quark operators using the method described above and
evaluating the integrals numerically instead of algebraically. In all
cases we find full agreement with the direct evaluation of the
diagrams. This independent calculation of the matching coefficients
provides a non-trivial check of the validity of our results.
In addition, we have also performed the calculation for the HYP
actions used in \cite{Lee:2003sk} and reproduce the result for the
matching matrix given in that reference.

\section{Tadpole improvement\label{sec:TI}}

Quite generally one finds that perturbative corrections in lattice
regularization are larger than those encountered in continuum
calculations, e.g. in the ${\overline {\rm MS}}$ scheme. The large
corrections typically arise from the additional diagrams that would be
absent in the continuum, in our case the fourth and fifth diagram in
Fig.~\ref{fig:graphs}. Since the size of the corrections is scheme
dependent they are not by themselves meaningful. Instead of just
working with the bare lattice parameters one can try to absorb the
large perturbative corrections into a redefinition of the basic
parameters of the theory, such as coupling constant, quark masses,
Wilson coefficients of the weak Hamiltonian, etc. Such techniques are
also used in the continuum, for example in the context of Heavy-Quark
Effective theories: the perturbative results for inclusive heavy hadron decay
rates get large perturbative corrections if they are expressed in
terms of the heavy quark pole mass but the corrections are much
smaller if one uses a more suitable mass definition
\cite{Bigi:1997fj,Czarnecki:1997sz,Hoang:1998ng}.  Similar ideas are
behind the so-called ``physical'' couplings and optimal scale setting
prescriptions, introduced quite some time ago \cite{Brodsky:1982gc}.

The large effects of the tadpole diagrams can be canceled by tadpole
improvement \cite{Lepage:1992xa}. Tadpole improvement is achieved by
working with an action in which all gauge links are divided by an
average link $u_{0}$. In the simulation the average link is determined
numerically, while it is evaluated perturbatively 
in the matching calculation. We define
\begin{equation}
u_0=1-\alpha_s\,u_0^{(1)}+\dots.
\end{equation}
 The average link receives large
perturbative corrections which then cancel out the large contributions
from the tadpole diagrams. In this section we give the tadpole
improvement terms relevant for our calculation. Some care is required
because different variants of tadpole improvement are used in the
literature. First of all, two different definitions of the the  mean link $u_0$ are common: the mean link in Landau gauge
\begin{equation}
u_0 =\left\langle \frac{1}{3}\,{\rm Re}\,{\rm Tr}\,U_{n,\mu} \right\rangle_{\xi=-1}\label{eq:meanLink}
\end{equation}
or the fourth root of the average plaquette
\begin{equation}
u_0^P=\left\langle \frac{1}{3}\,{\rm Re}\,{\rm Tr}\, U_{n,\mu}\,U_{n+\mu,\nu}\,U^{\dagger}_{n+\mu+\nu,\mu}\,U^{\dagger}_{n+\nu,\nu} \right\rangle^{1/4}\,.
\end{equation}
The second definition is somewhat simpler since no gauge fixing is
required and it is what MILC is using in their simulations with the
Asqtad action. Furthermore, MILC divides the $L$-link terms in the
action and operators not by $u_0^L$ as \cite{Lepage:1992xa} but by
$u_0^{L-1}$ and absorbs a power of $u_0$ into the fermion mass. With
their prescription no tadpole improvement factor is present in the
fermion part of the naive unimproved action
\cite{Orginos:1999cr}.\footnote{We thank C.~Bernard for pointing this
  out to us.} Since the fermion mass does not enter our calculation,
the difference between the two prescriptions merely amount to a
rescaling of the fermion field and leads to identical results for the
matching coefficients we calculate.

If we divide the $L$-link terms by $u_0^L$, the tadpole improvement
contribution from the current operators is
\begin{equation}
q^{({\rm tad})}=4\pi u_0^{(1)}\left(
\begin{array}{cccc}
6 & 0  & 0 & 0 \\
0 & 4 &  0 & 0  \\
0 & 0  & 2 & 0 \\
0 & 0 &  0 & 4  
\end{array}\right)\,.
\label{eqs:tadsop}
\end{equation}
In addition, there is a tadpole improvement correction to the square
of the wave function renormalization
\begin{align}
\delta Z^{\rm tad}_{\rm unimp.}\,Q_i&=-8\pi u_0^{(1)}\,Q_i\,, \nonumber \\
\delta Z^{\rm tad}_{\rm Asqtad}\,Q_i&=-18\pi u_0^{(1)}\,Q_i\,.
\label{eqs:z2}
\end{align}

The results given in the next section are obtained with $u_0$ defined
as the mean link in Landau gauge (\ref{eq:meanLink}). However, using
the explicit form of the tadpole improvement terms given above the
prescription can easily be changed.

\section{Wilson coefficients of the lattice operators\label{sec:results}}

The Wilson coefficients of the lattice operators can now be read off
by imposing that the four-quark scattering amplitudes are equal in the
continuum and on the lattice. Clearly, the solution to the matching
equation (\ref{eq:matching}) is in general not unique since there are
many different discretizations of the same continuum operator. In the
present case, the two $(\bar s d)_{V-A}\,(\bar s d)_{V-A}$ operators
with different color structure become equivalent in the continuum
limit where they are related by a Fierz transformation. The matching
condition will thus not fix the Wilson coefficients of the operators
with different color structure individually. To be specific, at tree level
the matching conditions impose only three relations
\begin{align}
C_{1}^{{\rm latt}}&=C_{3}^{{\rm latt}}+{\mathcal O}(\alpha_s) \,, \nonumber\\
C_2^{{\rm latt}}&=C^{{\rm latt}}_4+{\mathcal O}(\alpha_s) \,, \\
C^{{\rm cont}}&=C_{1}^{{\rm latt}}+C_{2}^{{\rm latt}} +{\mathcal O}(\alpha_s)\,,\nonumber
\end{align} 
among the four lattice operator Wilson
coefficients. Here, $C^{{\rm cont}}$ is the Wilson coefficient of the
continuum operator $Q$, while $C_i^{{\rm latt}}$ are the coefficients
of $Q_i$.

In lattice simulations, the quarks in the operator $Q$ are treated as
four distinct flavors \cite{Sharpe:1986xu}. In this case, there is no
ambiguity: the effective continuum Hamiltonian contains operators
$Q_i^{\rm cont}$ that are the continuum limit of each $Q_i$. Also,
instead of writing relations among coefficients, it is more convenient
to rewrite the result of the matching in the form 
\begin{equation} 
Q_i^{\rm cont}\equiv \sum_{j=1}^4\,C_{ij} Q_j^{\rm lat}, 
\end{equation} 
where ``$\equiv$'' means equality at the level of matrix elements, see
(\ref{eq:matching}). The 4 by 4 matrix $C_{ij}$ has a perturbative
expansion
\begin{equation}
C_{ij}=\delta_{ij}+\left ( \frac{\alpha_s}{\pi} \right )
\,C_{ij}^{(1)}+\dots 
\end{equation}
and is obtained from the results for the graphs in the previous section by
\begin{align}
C_{ij}^{(1)} &=-\frac{1}{4}\left[{\delta Z}^{(1)} \delta_{ij}+
\left (\frac{a\mu}{2} \right )^{2\epsilon}\sum \limits_{\alpha} 
q_{\alpha,j}^{i} + \delta Z^{(1)}_{ij}\right] \, . \label{match}
\end{align}
Here $\delta Z^{(1)}$ stands for the wave function 
renormalization and  $Z_{ij}=\delta_{ij}+\frac{\alpha_s}{\pi} Z^{(1)}_{ij}$ 
is the matrix of the ${\overline {\rm MS}}$ renormalization constants 
 for the operators
$Q^{\rm cont}_i$ in the continuum. For the purposes of 
computing $C_{ij}^{(1)}$, the role of this matrix 
 is to remove any residual $1/\epsilon$ dependence in (\ref{match}).
Using the results for the matrices $q_{\alpha}$ 
given in the previous section, it is straightforward
to compute the matching coefficients. We split the matrix $C_{ij}^{(1)}$ into
 three parts
\be
C^{(1)} = \ln \left (\frac{a\mu}{2} \right )\,\gamma^{(1)} + C^{(1)}_C+C^{(1)}_U.
\ee
Here, $\gamma^{(1)}$ is the universal one-loop anomalous dimensions matrix of the operators $Q_i$:
\be
\gamma^{(1)} = \left ( 
\begin{array}{cccc}
0 & 0  & \frac{1}{2} & -\frac{3}{2}   \\ 
-\frac{3}{4}& \frac{9}{4}& -\frac{3}{4}& -\frac{7}{4} \\
\frac{1}{2}& -\frac{3}{2}& 0& 0 \\ 
-\frac{3}{4}& -\frac{7}{4}& -\frac{3}{4}& \frac{9}{4}
\end{array}
\right ).  
\ee

The remainder $C_{C}+C_{U}$ depends on the lattice action and we give
it for three cases: (a) Asqtad fermion and improved gauge action, (b)
Asqtad fermion with plaquette gluon action, (c) standard action, no
improvement. The reason for splitting the remainder into two pieces
which are not separately gauge invariant, is as follows. Occasionally,
simulations are performed with operators that are obtained by dropping
the links that connect fermions on different sites. The simulations
with these gauge non-invariant operators are performed in Landau gauge
and their matching coefficients can be extracted from our calculation.
To obtain them, we need to account for the contributions from graphs
$1-3$ in Fig.~\ref{fig:graphs}, the wave function renormalization
contributions as well as the tadpole improvement for the wave function
given in (\ref{eqs:z2}).  We denote the corresponding matching matrix
by $C_{C}$.  The remainder, denoted by $C_{U}$ is the contribution
from graphs 4 and 5 which arises from the link fields in the current
operator and the operator tadpole improvement term given in
(\ref{eqs:tadsop}).  We give the matrices $C_{C,U}$ in Landau gauge;
their sum is gauge invariant.

For the Asqtad and improved gauge action, we
find
\begin{align}
C_{C}^{(1a)} &= \left (
\begin{array}{cccc}
2.83(14) & -0.75 & 0.38(1) & -1.15(2)  \\ 
-1.25(1) & 4.33(10) & -0.58(1) & -1.34(3)  \\ 
0.38(1) & -1.15(2) & 2.59(15) & -0.75  \\ 
-0.58(1) & -1.34(3) & -1.34(2) & 4.36(10)
\end{array}
\right ), \nonumber\\ & \\
C_{U}^{(1a)} &= \left (
\begin{array}{cccc}
2.08(5) & 0 & 0 & 0  \\ 
0.29 & 0.98(3) & 0 & 0  \\ 
0 & 0 & 0 & 0  \\ 
0 & 0 & -0.20(1) & 1.14(3) 
\end{array}
\right ).\nonumber
\end{align} 
These results are obtained after tadpole improvement with the ``mean
link in Landau gauge'' definition of the improvement parameter, 
$u_0^{(1)}=0.750(2)$ for the improved gluon action and 
$u_0^{(1)} = 0.97432$ for the unimproved action. For Asqtad with the standard
plaquette gluon action, the result is
\begin{align}
C_{C}^{(1b)} &= \left (
\begin{array}{cccc}
2.33 & -0.75 & 0.34 & -1.03  \\ 
-1.18 & 3.61(1) & -0.51 & -1.20  \\ 
0.34 & -1.03 & 2.09 & -0.75  \\ 
-0.51 & -1.20 & -1.27 & 3.64(1)
\end{array}
\right ), \nonumber\\ & \\
C_{U}^{(1b)} &= \left (
\begin{array}{cccc}
2.97 & 0 & 0 & 0  \\ 
0.32 & 1.43 & 0 & 0  \\ 
0 & 0 & 0 & 0  \\ 
0 & 0 & -0.20 & 1.60  \\ 
\end{array}
\right ).\nonumber
\end{align}
Notice that case (b) is not equivalent to what is considered in
\cite{Lee:2003sk}. These authors employ a different discretization of
the operators using insertions of fattened links to make the operators
gauge invariant while we are using thin links. Also, their fermion
action is similar but not identical to the fermion part of the Asqtad
action. 

Without any improvement terms in the action, we get
\begin{align}
C_{C}^{(1c)} &= \left (
\begin{array}{cccc}
-0.81 & -0.75 & 0.18 & -0.54  \\ 
-0.85 & -0.5 & -0.27 & -0.63  \\ 
0.18 & -0.54 & 0.25 & -0.75  \\ 
-0.27 & -0.63 & -0.46 & -0.63 
\end{array} \right ) ,  \nonumber\\
& \\
C_{U}^{(1c)} &= \left (
\begin{array}{cccc}
3.83 & 0 & 0 & 0  \\ 
0.47 & 1.86 & 0 & 0  \\ 
0 & 0 & 0 & 0  \\ 
0 & 0 & -0.38 & 2.14
\end{array} \right ) .  \nonumber
\end{align}
The matrices for the unimproved action agree with
the known results in the literature \cite{Ishizuka:1993fs,Lee:2003sk}.

We write the finite correction to the normalization of the
axial-vector current in the form $Z_A=1+\frac{\alpha_s}{\pi} \delta
Z_A$ and obtain
\begin{align}
  \delta Z_A^{(a)}&=1.17(7), & \delta Z_A^{(b)}&=0.921, & \delta
  Z_A^{(c)}&=0.
\end{align}
The current renormalization happens to be identical for the gauge
invariant and non-invariant operators in Landau gauge. This occurs
because in this gauge, the contribution of the fourth graph vanishes
and the contribution from fifth graph gets canceled by tadpole
improvement.

\section{Summary and discussion}

We have calculated the matching of the Wilson coefficients in the
$\Delta S=2$ flavor changing effective Hamiltonian from the continuum
to the lattice for the Asqtad lattice action. Together with
non-perturbative calculations of the $K-\bar K$ matrix element of
the lattice operators, our results determine the amount of indirect CP
violation in the kaon system. The matrix elements are currently being
evaluated by the HPQCD and UKQCD collaborations \cite{Gamiz:2004qx,CDGSW05}.  
The calculation uses the configurations with $n_f=2+1$ dynamical flavors  
generated by the MILC collaboration. 

Previous calculations of the quantity $B_K$, which determines the
strength of $K-\bar K$ mixing, were performed in the quenched
approximation. This induces an essentially unknown and
irreducible systematic error into the result. Precise simulations with
{\em dynamical} fermions are necessary in order to be able to make
full use of the experimental data on $K-\bar K$ mixing to constrain
the CKM matrix elements entering the effective Hamiltonian. Among the
currently available fermion actions, only staggered actions allow for
simulations with low statistical errors at small quark masses.
However, the unimproved staggered action suffers from large
taste-changing interactions. The Asqtad action, for which we have performed
the matching, has been designed to reduce these as well as other
cut-off effects.

We have performed our calculation with two independent methods. First,
we have evaluated the diagrams in two different ways: by directly
calculating the various diagrams and by first separating off the part
which can be inferred from the renormalization of the current
operators. Second, we have evaluated the lattice integrals both
algebraically and numerically.  For the algebraic evaluation, we have
expanded the diagrams around the continuum limit.  This produces a set
of lattice tadpole integrals which we then reduced to a minimal set of
master integrals using computer algebra. The agreement between these
two rather different methods provides a strong check on our results.

We find that the size of one-loop corrections to the tree level
matching for the Asqtad action is very similar to what is found with
unimproved staggered and other improved staggered actions, such as the
HYP action \cite{Lee:2004qv}. In quenched calculations at a lattice
spacing $1/a=1.6~{\rm GeV}$, the shift in the value of the bare $B_K$
due to ${\cal O}(\alpha_s)$ corrections is around 10\% for all these
actions \cite{CDGSW05}.  In dynamical simulations with the Asqtad
action the shift turns out to be over 15\% at the same lattice
spacing, simply because the value of the strong coupling is much
larger for $n_f=3$ than for $n_f=0$ dynamical flavors. Also, it turns
out that the values for the matching coefficients are not very
sensitive to the improvement of the gluon action. Switching off the
gluon improvement terms in the Asqtad action changes the entries of
the one-loop matching matrix by less than 20\% and the changes are
even smaller for the larger elements on the diagonal. The motivation for
using an improved staggered action in the calculation of $B_K$ is not
to reduce the size of the one-loop corrections, which are already small
in the unimproved case, but to correct the large ${\cal O}(a^2)$
scaling violations that highly affect the unimproved case; see
\cite{Gamiz:2004qx,CDGSW05} for further discussion.

The situation is different for the calculation of other weak matrix
elements relevant in the study of CP-violating effects in the
kaon system. For the unimproved staggered action, Sharpe and Patel
\cite{Sharpe:1993ur} found very large perturbative corrections for
operators with scalar and tensor Dirac structure. The contributions
are large even after tadpole improvement, which casts doubt on the
perturbative expansion for the unimproved action. The corrections are
reduced to an acceptable level with the HYP action
\cite{Hasenfratz:2001hp} and operators which are made gauge invariant
by the insertion of fattened links \cite{LS02, Lee:2003sk}. In
order to check whether these large corrections are present for the
Asqtad action with standard four-quark operators, we have analyzed the
mixing of scalar operators.  We find that the corrections are small
with the improved action. The suppression arises because the
improvement terms reduce taste-changing interactions. In Appendix
\ref{sec:scalar} we present the results for the mixing of scalar
operators and discuss the suppression of the large contributions in
more detail.

The absence of anomalously large one-loop corrections suggests that
the two-loop corrections to our results should be reasonably small.
However, in order to reduce the uncertainty in dynamical calculations
to the level of a few per-cent, a two-loop matching calculation will
be necessary.

{\bf Acknowledgments} We are grateful to Christine Davies and Andreas
Kronfeld for discussions and comments on the manuscript. We thank the
KITP in Santa Barbara and the INT in Seattle where part of this work
was carried out for their hospitality and financial support. This
research was supported in part by the U.S.\ Department of Energy
contracts DE-AC02-76CH03000, DE-FG03-94ER-40833 and the Outstanding
Junior Investigator Award DE-FG03-94ER-40833, and by the Alfred P.
Sloan Foundation. E.G.\ is indebted to the European Commission for a
Marie-Curie Grant No.\ MEIF-CT-2003-501309. Fermilab is operated by
Universities Research Association Inc., under contract with the U.S.\
Department of Energy.

\appendix

\section{Mixing of scalar operators\label{sec:scalar}}

In this appendix we evaluate the matching for the operators
\begin{equation}
\begin{aligned}\label{scaloperators}  
Q^S_{1} &= (\bar \psi_1^a \psi_2^a)_S (\bar \psi_3^b \psi_4^b)_S, \\
 Q^S_{2}&= (\bar \psi_1^a \psi_2^b)_S (\bar \psi_3^b \psi_4^a)_S,
\end{aligned}
\end{equation}
where the bilinears $(\bar \psi_i^a \psi_j^a)_S$ have unit Dirac and
taste structure, $\gamma_S\otimes\xi_T=1\otimes 1$ (or equivalently
$\gamma_5\otimes \gamma_5$). These operators are not needed for the
evaluation of $B_K$, but scalar and pseudoscalar operators are present
in the $\Delta S=1$ effective Lagrangian used to determine
$\varepsilon'$. The scalar operators mix into tensor operators under
renormalization, but the form of this mixing is not important for our
discussion.

For gauge non-invariant operators, the above matching has been studied
by Sharpe and Patel \cite{Sharpe:1993ur}, who find very large
perturbative corrections for these operators even after tadpole
improvement. Such large corrections are absent for the HYP action
\cite{Hasenfratz:2001hp} with operators which are made gauge invariant
by the insertion of fattened links \cite{LS02, Lee:2003sk}. As we show
below, this is true also for the Asqtad action with standard operators.

Using the same notations and conventions as in
Section \ref{sec:results}, the $2\times 2$ matrix for the anomalous
dimension of the operators in (\ref{scaloperators}) is
\begin{equation}
\gamma^{(1)}=\left(\begin{matrix} 4 & 0 \\ \frac{3}{2} & -
    \frac{1}{2}\end{matrix}\right)\,. 
\end{equation}
After tadpole improvement, the corresponding matching matrices for 
the unimproved action are
\begin{align}\label{eq:scalarunimproved}
C_{C}^{(1c)} &= \left (
\begin{array}{cccc}
-17.26 & 0 \\ -6.28 & 1.59 
\end{array} \right ) ,  \nonumber \\
& \\
C_{U}^{(1c)} &= \left (
\begin{array}{cccc}
0 & 0 \\ -2.45 & 2.83
\end{array} \right ) .  \nonumber
\end{align}
The one-loop correction to the matching is indeed very large, in
particular for the element that describes the mixing of $Q^S_{1}$ into
itself. The large correction is entirely due to the first diagram in
Fig.~\ref{fig:graphs}. It is twice the renormalization of the scalar
current. Note that the operator $Q^S_1$ is completely local. In
terms of the naive field, it is simply
\begin{equation}
 \bar q\, (1\otimes 1)\, q
= \frac{1}{16}\sum_{A} \bar\psi(A)\,\psi(A) \,.
\end{equation}
This implies that for the unimproved action large values are obtained,
independently of the prescription adopted to make the operators gauge
invariant. 

The values obtained for the Asqtad action are
\begin{align}
C_{C}^{(1a)} &= \left (
\begin{array}{cccc}
-1.8(1) & 0 \\ -0.6(1) & 3.6(1)
\end{array} \right ) ,  \nonumber \\
&\\
C_{U}^{(1a)} &= \left (
\begin{array}{cccc}
0 & 0 \\
-1.81(1) & 1.68(3) 
\end{array} \right ) .  \nonumber
\end{align}
It is comforting to see that the corrections are smaller with the
improved action. The dramatic reduction is mostly due to the fat-link
terms in the fermion action.

The origin of the large corrections for the unimproved action, such as
those in (\ref{eq:scalarunimproved}), has been analyzed by Golterman
\cite{Golterman:1998jj}. He splits the integration region for the loop
momenta $\pi< k_\mu \leq \pi$ into a region around zero
$\pi/2<k_\mu\leq\pi/2$ and a remainder which contains the corners of
the Brillouin zone. For Wilson fermions, he finds that after tadpole
improvement the main contribution arises from the integration region
around zero. For staggered fermions, on the other hand, large
corrections arise from the corners of the Brillouin zone. In this
region, the gluon propagator is off-shell, but the staggered fermion
propagator has poles.\footnote{Note that the contribution we are
  talking about is from the region of hard loop momentum; there are no
  singularities from propagating doublers. Such contributions only
  arise at higher order in the expansion in the lattice spacing.}
Because the gluon propagator is far off-shell, the first graph in
Fig.~\ref{fig:bilgraphs} can be viewed as a fermion tadpole in the
corner region. This is the interpretation put forward by Golterman.

The reason for the reduction achieved with the improved action is that
the improved quark gluon coupling is designed to suppress
taste-changing interactions. The quark-gluon interaction switches off
if the in- and outgoing tastes differ. This mechanism
suppresses the contribution from the corners of the Brillouin zone to
the first diagram in Fig.~\ref{fig:graphs}, which corresponds
precisely to the situation where the taste of the internal quark line
is different from the taste of the external line.

To conclude, we find that the anomalously large perturbative
corrections present in some cases for the unimproved staggered action
are suppressed in the Asqtad results due to the smaller taste-symmetry 
breaking of this action. We thus expect that the two-loop corrections to our
matching calculation will be reasonably small.

%\newpage

\end{document}